\documentclass[final,1p,times]{elsarticle}

\usepackage{graphics}
\usepackage{amssymb}
\usepackage{color, colortbl}
\usepackage{lineno}
\biboptions{comma,sort&compress}
\definecolor{Gray}{gray}{0.9}
\definecolor{honey1}{rgb}{0.94,1,0.94}
\definecolor{honey2}{rgb}{0.75,0.8,0.75}
\usepackage{pdflscape}
\usepackage{everypage}
\usepackage{lipsum}
\usepackage{supertabular}

 \usepackage{color,soul}
\newcommand{\Lpagenumber}{\ifdim\textwidth=\linewidth\else\bgroup
  \dimendef\margin=0 
  \ifodd\value{page}\margin=\oddsidemargin
  \else\margin=\evensidemargin
  \fi
  \raisebox{\dimexpr -\topmargin-\headheight-\headsep-0.5\linewidth}[0pt][0pt]{%
    \rlap{\hspace{\dimexpr \margin+\textheight+\footskip}%
    \llap{\rotatebox{90}{\thepage}}}}%
\egroup\fi}
\AddEverypageHook{\Lpagenumber}%

\journal{Future Generation Computer Systems}

\begin{document}

\begin{frontmatter}


\title{The Role of Big Data Analytics in Industrial Internet of Things}
\author[label1]{Muhammad Habib ur Rehman}
\author[label2]{Ibrar Yaqoob}
\author[label3]{Khaled Salah}
\author[label4]{Muhammad Imran}
\author[label5]{\newline Prem Prakash Jayaraman}
\author[label6]{Charith Perera}

\address[label1]{National University of Computer and Emerging Sciences, Lahore, Pakistan}
\address[label2]{Department of Computer Science and Engineering, Kyung Hee University, Yongin-si 17104, South Korea}
\address[label3]{Department of Electrical and Computer Engineering, Khalifa University, UAE}
\address[label4]{College of Computer and Information Sciences, King Saud University, Riyadh, Saudi Arabia}
\address[label5]{Department of CS \& SE, Swinburne University of Technology, VIC 3122, Australia}
\address[label6]{School of Computer Science and Informatics, Cardiff University, United Kingdom}

\begin{abstract}
Big data production in industrial Internet of Things (IIoT) is evident due to the massive deployment of sensors and Internet of Things (IoT) devices. However, big data processing is challenging due to limited computational, networking and storage resources at IoT device-end. Big data analytics (BDA) is expected to provide operational- and customer-level intelligence in IIoT systems. Although numerous studies on IIoT and BDA exist, only a few studies have explored the convergence of the two paradigms. In this study, we investigate the recent BDA technologies, algorithms and techniques that can lead to the development of intelligent IIoT systems. We devise a taxonomy by classifying and categorising the literature on the basis of important parameters (e.g. data sources, analytics tools, analytics techniques, requirements, industrial analytics applications and analytics types). We present the frameworks and case studies of the various enterprises that have benefited from BDA. We also enumerate the considerable opportunities introduced by BDA in IIoT. We identify and discuss the indispensable challenges that remain to be addressed as future research directions as well.
\end{abstract}


\end{frontmatter}


\section{Introduction}
\label{S:1}
Industrial Internet of Things (IIoT) (also known as Industry 4.0), which was initially conceived as a vision by the German government, is currently attributed as the fourth industrial revolution. The technology ecosystem underpinning IIoT is mainly the integration of cyber–physical systems (CPS)~\cite{akkaya2016systems}, Internet of Things (IoT), cloud computing~\cite{wollschlaeger2017future,perera2017valorising,perera2014survey}, automation (e.g. intelligent robots in product assembly lines)~\cite{pedersen2016robot}, Internet of services~\cite{liu2016services}, wireless technologies, augmented reality~\cite{kim2016augmented} and concentric computing~\cite{concentric2017}, amongst others. Advances in such related areas as IoT, big data analytics (BDA), cloud computing and CPS have fuelled the formation of IIoT activities to deliver unprecedented flexibility, precision and efficiency to manufacturing processes~\cite{yaqoob2017internet, ahmed2016internet}. Given this cross-platform integration, IIoT systems need to ensure interoperability, virtualisation, decentralisation, real-time capability, service orientation, modularity and security across all verticals~\cite{wang2016intelligent}. However, these systems are perceived to have qualities, such as self-awareness, self-prediction, self-comparison, self-configuration, self-maintenance and self-organisation~\cite{lee2015smart}.

BDA is a related area that enables IIoT systems to deliver value for data captured from cross-platform integration. BDA refers to the process of collecting, managing, processing, analysing and visualising continuously evolving data in terms of volume, velocity, value, variety and veracity~\cite{marjani2017big}. Big data in IIoT systems arise due to unbounded internal and external activities relevant to customers, business operations, production and machines~\cite{ur2016big}. BDA processes in IIoT systems manage the collected data using multiple transient and persistent storage systems that provide on-board, in-memory, in-network and large-scale distributed storage facilities across IIoT systems~\cite{geng2016research,gaberinternet}. The granularity of data processing facilities for BDA processes in IIoT systems vary from resource-constrained IoT devices to resourceful large-scale distributed cloud computing systems~\cite{ur2016towards}. Similarly, analytic operations differ in terms of descriptive, prescriptive, predictive and preventive procedures~\cite{ur2016big}. In addition, BDA processes must ensure real-time knowledge visualisation across multiple IIoT systems. A proper integration of BDA processes into IIoT systems is perceived to maximise value creation to evolve business models for profit maximisation~\cite{ehret2017unlocking,ur2016big}.

\subsection{Motivation}
Although IIoT~\cite{da2014internet, mumtaz2017massive, al2018context, chaudhary2018sdn, perera2015context, perera2015emerging} and BDA~\cite{ahmed2017role, marjani2017big, Tsai2015, 9876555, SAGGI2018758, wang2016big, hu2014toward, li2017big} have been widely studied separately, only a few studies including~\cite{ur2018big} have explored the convergence of the two domains.

Big data production in IIoT is evident due to large-scale deployment of sensing devices and systems in pervasive and ubiquitous industrial networks. Given that the concept of IIoT systems is still evolving, complete integration and implementation of BDA processes in IIoT systems are unavailable yet~\cite{radhakrishnan2016convergence, ur2018big}. Existing surveys on IIoT systems focus on concepts related to adoption of IIoTs~\cite{Georgakopoulos2016, lee2015internet}, the integration of IIoTs and edge cloud computing systems~ \cite{georgakopoulos2016internet}, industrial marketplaces for IIoTs~ \cite{perera2014survey}, big data and virtualisation technologies for IIoT systems~\cite{babiceanu2016big}, technological advancements relevant to CPS in IIoT systems~\cite{wang2015current}, smart manufacturing~\cite{kang2016smart} and big data applications for business operations~\cite{addo2016big,choi2017recent,chang2016mobile}. We introduced the concept of the concentric computing model (CCM) for BDA in IIoT in our previous work~\cite{ur2018big} whereby we outlined the discussion on different layers of CCM and discussed the relevant research challenges that must be addressed to fully enable CCM for BDA in IIoT. However, to the best of our knowledge, a detailed review on BDA implementation for IIoTs is still lacking in the existing literature. Thus, the current study presents the key operations of BDA for value creation in IIoT systems. On the basis of BDA concepts, this study surveys earlier contributions relevant to data analysis in IIoT systems.

\subsection{Contributions} 
The main contributions of this study are listed as follows.
\begin{itemize}
\item	We build a case of BDA for IIoT systems whereby the role and entire process of BDA are discussed. The study sets a theoretical ground to understand modern automated data pipelines for enriching intelligence in IIoT systems.
\item	We investigate existing state-of-the-art research studies on IIoT in terms of BDA. In this context, we categorise and classify the literature by devising a taxonomy.
\item	We present frameworks and case studies whereby BDA processes are adopted to improve the overall performance of IIoT systems.
\item	We present several research opportunities, challenges and future technologies to minimise the research gaps between state of the art (i.e. proposed in the literature) and state of the practice (i.e. adopted by industries in practice).
\end{itemize}

The rest of the paper is organised as follows. Section~\ref{bda} discusses the key concepts relevant to BDA in IIoT systems, followed by a detailed survey of existing technologies and algorithms in Section~\ref{survey}. Section~\ref{Taxo} presents the taxonomy, and Section~\ref{frameworkscase} highlights a few frameworks and relevant case studies. Section~\ref{future} presents the opportunities, open challenges and future directions. Section~\ref{conclusions} provides the concluding remarks. 

\section{BDA in IIoT Systems}
\label{bda}
This section presents a detailed discussion on different aspects of big data adoption in IIoT systems. To this end, several design principles, which should be considered prior to configuring and deploying IIoT systems, are highlighted. The role of BDA and its life cycle is discussed in detail to deliver end-to-end intelligence in IIoT systems.

\subsection{Design Principles for IIoT Systems}
\begin{figure}[h!]
	\centering
	\includegraphics[width=\textwidth]{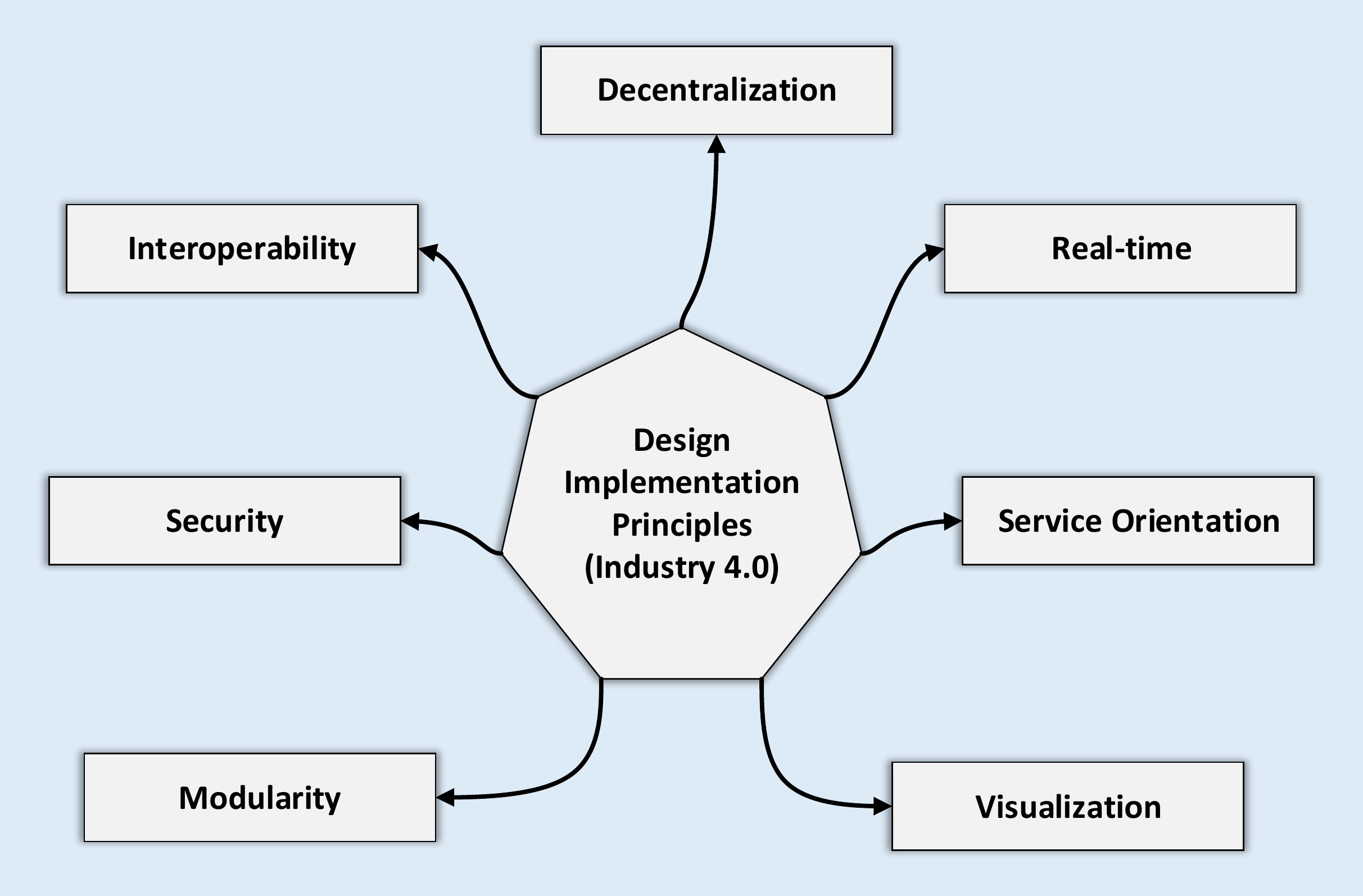}
	\caption{Seven design implementation principles for Industry 4.0 systems.}
	\label{des}
\end{figure}

The designs of IIoT systems involve seven principles~\cite{wang2016intelligent}, as depicted in Fig.~\ref{des}. Firstly, interoperability must be ensured amongst different technologies, such as CPS, IoT devices and concentric computing systems. Wireless data communication technologies play an unparalleled role to realise an interoperable system. Secondly, virtualisation technologies at all levels must be considered for efficient service provisioning and delivery across IIoT systems. Virtualisation varies in terms of platforms, networks, data, operating systems and applications. Thirdly, decentralisation must be conducted to ensure highly distributed IIoT systems. Decentralisation varies in terms of system-wide data processing and data storage. Fourthly, IIoT systems must provide real-time feedback to all stakeholders. Fifthly, service-orientation must be guaranteed whereby all system functions are implemented in the form of service-oriented architecture (SOA). Sixthly, modular approach must be adopted for system implementation. Lastly, system-wide security must be considered as core principle. The BDA process for IIoT systems must be designed in consideration of the above-mentioned principles.
 
 \begin{figure*}[b!]
 \centering
 \includegraphics[width=4.5in]{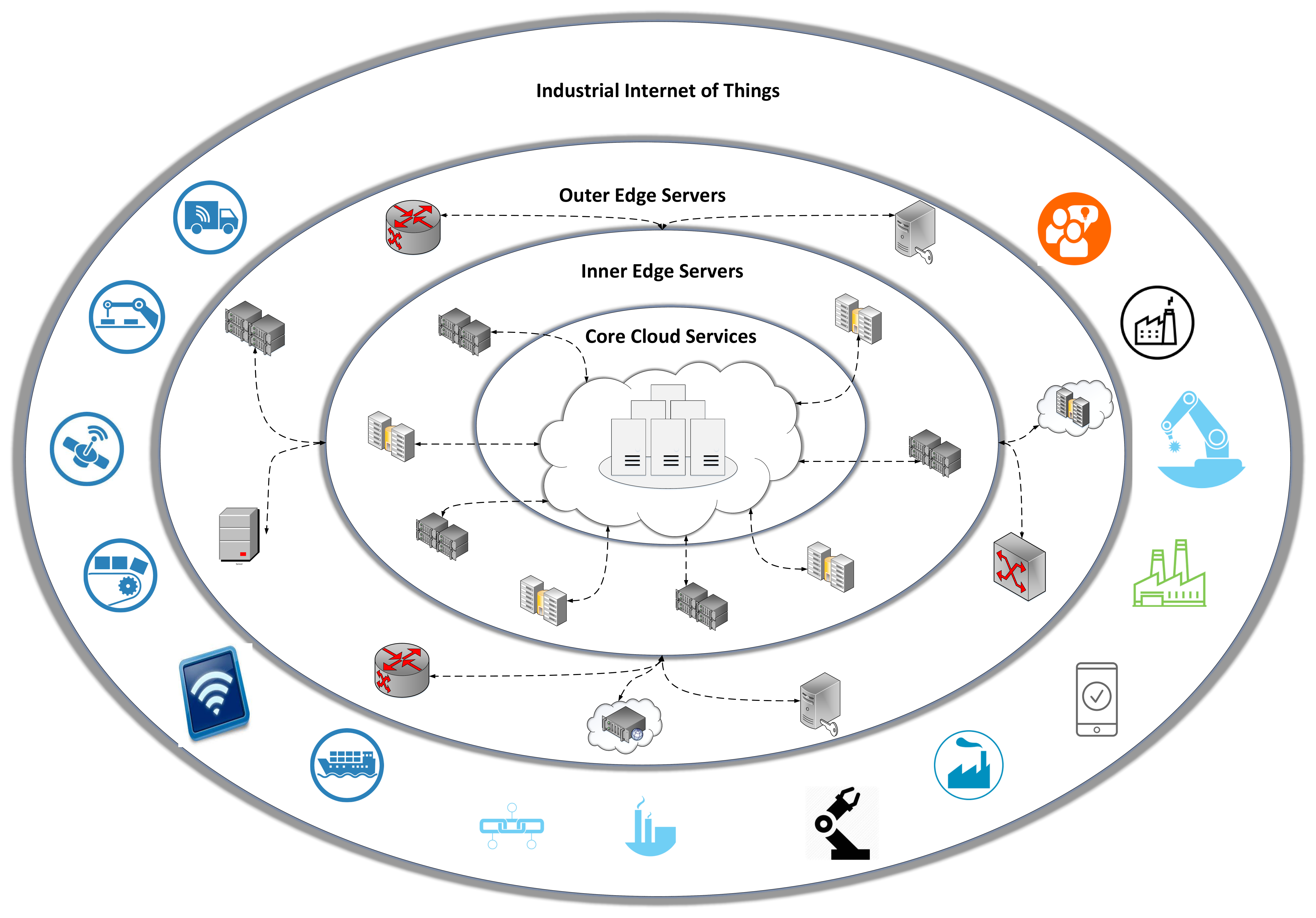}
 \caption{Industrial IoTs and Multilayer Computing Resources}
 \label{bdiagram}
 \end{figure*}
 
\subsection{Rise of Big Data in IIoT Systems}
Big data in IIoT systems emerge from a plethora of technologies. CPS refers to the integration of physical machine components with on-board computations and networking facilities~\cite{zhou2016cyber,wang2015current}. CPS and IoT devices act as the backbone of IIoT systems and thus generate massive amount of raw data streams, which result in big data~\cite{lee2015industrial}. Therefore, real-time analysis of these data can improve machine health and lead to defect-free product manufacturing~\cite{akkaya2016systems,Georgakopoulos2016,cheng2016industry}. 

IoT devices in IIoT systems refer to devices that can remotely sense and actuate in industrial environments~\cite{cheng2015study}. IoT devices either work as stand-alone devices that roam around industrial environments or are attached with existing CPS to perform certain predefined actions~\cite{zhong2017big}. The on-board sensing facilities in IoT devices lead the generation of big data, which may become useful for value creation in enterprises.
The integration of CPS and IoT devices results in massive back-end cloud service utilisation for the execution of BDA processes~\cite{tao2014cciot}. To achieve massively customised production, the number of cloud services can be grown immensely. Thus, BDA can facilitate in-service selection, service orchestration and real-time service provisioning~\cite{hossain2016cloud}. 

\subsection{Concentric Computing Model for BDA in IIoT}
Recent evolution in sensing and computing technologies has opened new avenues for big data processing. Concentric computing refers to the large-scale highly distributed computing systems based on a wide range of devices and computing facilities in different form factors~\cite{concentric2017}. Concentric computing offers big data processing at sensors levels, endpoints in IIoT systems, edge servers, and centralised and decentralised cloud computing systems, as illustrated in Fig.~\ref{bdiagram}~\cite{ur2016big,gupta2016ifogsim,georgakopoulos2016internet}. Despite their small size and limited computational power, sensors and IoT devices have the ability to filter and reduce raw data streams by using on-board smart data reduction strategies~\cite{ur2017emergence}. However, edge servers at gateways and centralised computing clusters have the ability to distribute the computing load for BDA applications \cite{lin2019data, pace2019edge}.
Multistage execution, automating, and management of BDA processes (i.e., data engineering, data preparation and data analytics) are necessary in concentric computing environments (such as sensors and wearable devices as endpoints, IoT devices, edge servers, and cloud computing servers)~\cite{8450541}.

\begin{figure*}[!t]
	\centering
	\includegraphics[scale=0.35]{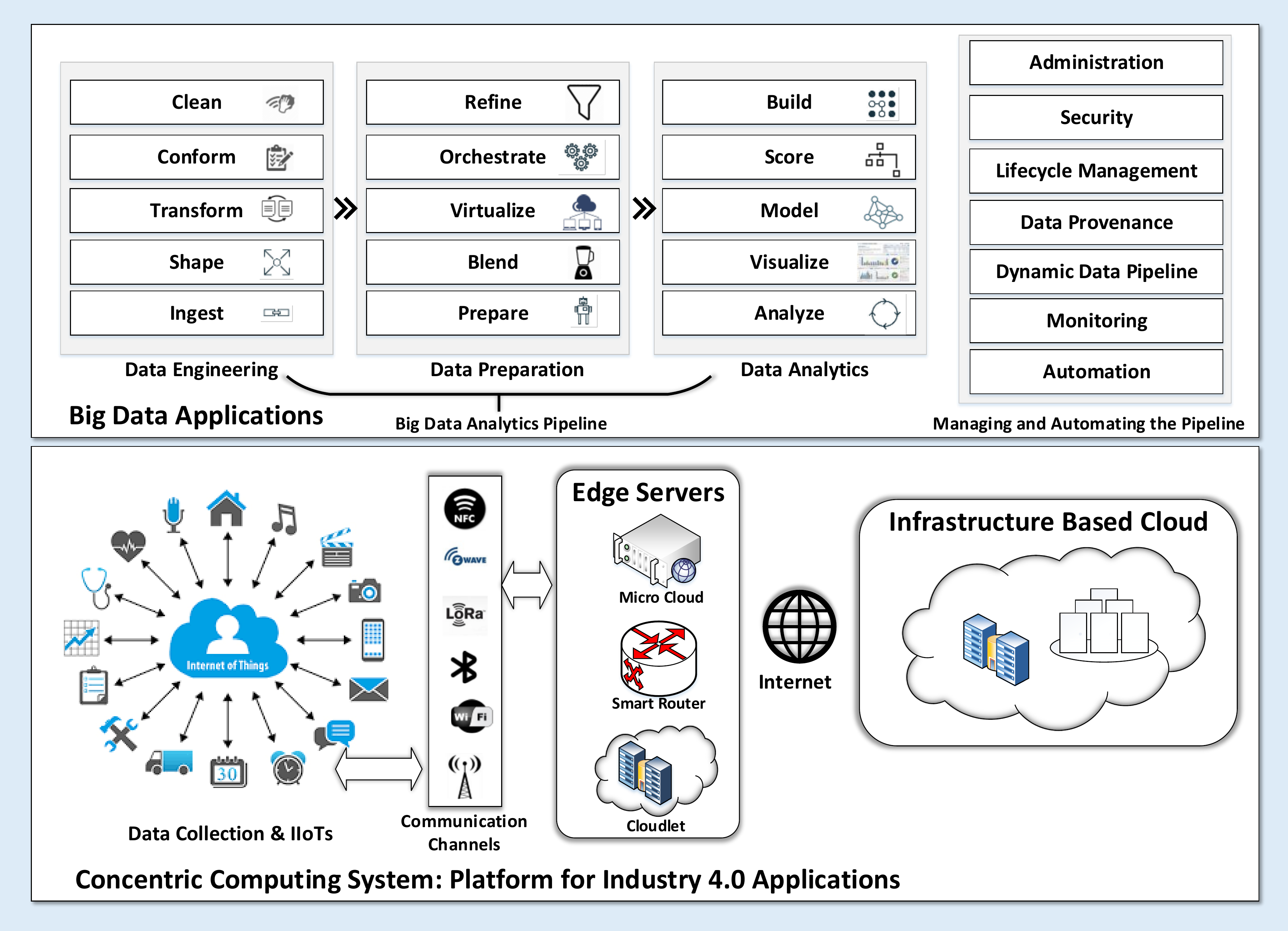}
\caption{Multistage execution, automating, and management of BDA processes (i.e., data engineering, data preparation and data analytics) in concentric computing environments (such as sensors and wearable devices as endpoints, IoT devices, edge servers, and cloud computing servers)~\cite{ur2018big}.}
\label{BDA_CCM}
\end{figure*}

\subsection{Big Data Analytics for Delivering Intelligence in IIoT Systems}
BDA processes are executed as a result of multistage highly interdependent application components (Fig.~\ref{BDA_CCM}). These components are categorised as follows.

\subsubsection{Data Engineering}
Data engineers build computing and storage infrastructure to ingest~\cite{ji2016device}, clean~\cite{wang2016cleanix}, conform~\cite{haryadi2016antecedents}, shape~\cite{lefevre2016building} and transform~\cite{wang2017integrated} data. IIoT systems produce and ingest big data from inbound enterprise operations and outbound customer activities. The raw data at the earliest stage need further processing to improve the quality and establish the relevance with IIoT applications. Therefore, data wrangling and cleaning methodologies help select relevant datasets in case of historical data or data streams in case of streaming data. Data conformity procedures are applied to ensure relevant, correctly collected big data. Data shaping and transformation methodologies help improve data quality by reducing the number of attributes and converting data formats for uniform data processing. 

\subsubsection{Data Preparation}
Big data emerge in raw form with large volume and enormous speed, and data scientists spend $70\%-80\%$ of their time in data preparation activities~\cite{cleaning}. Big data are refined using statistical methods to handle unstructured, unbalanced and nonstandardised data points efficiently~\cite{wiener2016biggis}. In addition, data refinement helps summarise voluminous data to reduce overall complexity. As a result, the spatiotemporal attributes of big data in IIoT systems vary. Ultimately, data locality is necessary to reduce in-network traffic and latency in big-data applications~\cite{wiener2016biggis}. Location-aware highly virtualised data infrastructure can address these issues. However, data blending, which is the process of combining data from multiple sources, becomes complex. Accordingly, further involvement by data scientists~\cite{ong2016data} is required to perform data cleaning and noise removal~\cite{garcia2017enabling}. Detection methods for outliers and anomalies are also needed to prepare big data for further analysis~ \cite{bai2016efficient,stojanovic2016big}.

\subsubsection{Data Analytics}
The analytic processes in IIoT systems are executed in multiple phases~\cite{nino2016requirements}. Data scientists generate learning models from high-quality well-prepared data. After the model is developed, model scoring operations are performed by giving sample datasets and finding and ranking the attributes in datasets/data streams. The correctly tuned models are deployed in production environments to find the knowledge patterns from future data. 

\subsubsection{Managing and Automating the Data Pipeline} 
Although existing literature still lacks the concept of automated data pipelines in IIoT systems, BDA processes are executed as a sequence of operations during data engineering, preparation and analytics. Therefore, a holistic approach is needed to execute and administer BDA processes across all layers of concentric computing systems. Life cycle management is needed for full process execution from raw data acquisition to knowledge visualisation and actuation. Data provenance, that is, designating ownership of data to different stack holders, also needs serious attention to ensure system-wide control on data~\cite{wang2016logprov}. The continuous evolution in data streams results in knowledge shift that enforces data pipelines to adaptively reconfigure analytic processes. The data pipelines need to be continuously monitored for change detection, and the entire BDA process needs to be re-executed to produce high-quality results~\cite{yamanishi2016detecting}. In security perspective, the cross-platform execution of BDA processes demands secure operations at IoT device, CPS and big data levels~\cite{cheng2017exploiting}.

\section{Technologies and Algorithms for BDA in IIoT systems} 
\label{survey}
A common example of IIoT systems is the concept of a smart factory system (SFS)~\cite{georgakopoulos2016internet}. The key attributes of SFS and its subsystems are self-awareness, self-organisation, self-maintenance, self-prediction, self-configuration and self-comparison~\cite{lee2015smart}. This section presents the review of early studies that presented BDA in the context of SFS and IIoT systems~\cite{lee2015smart} in consideration of the aforementioned autonomy related attributes (Table~\ref{tabletechnologies}).
\subsection{Mass Product Customization towards IIoT Lean Manufacturing}
Although the main objective of IIoT systems is to maximise production considering massive customisation in accordance with customer requirements, the existing literature still lacks an end-to-end predictive analytics framework. Computational intelligence-based methods, such as self-organising map (SOM) algorithms, are used to optimise big data for feeding in the production systems and enable massively customised product manufacturing~\cite{saldivar2016self}. The neural network-based SOM algorithm effectively enables smart production cycle in SFS. The cycle is based on a close loop within a sequence of operations, including smart design, manufacturing, production and services whereby feedback is collected after each cycle and subsequent operations at each stage are improved. Clustering-based big data optimisation is another approach whereby k-means clustering algorithms are used to cluster the attributes from customer data. The produced clusters are used to intelligently improve the design process in the product life cycle~\cite{saldivar2016identifying}. Another alternate for massive product customisation is the adoption of cloud-based manufacturing systems whereby big data integration is performed in cloud computing systems~\cite{yang2017towards}. However, the resultant big data are integrated from multiple sources, such as social media data streams relevant to customer behaviour and IIoT data streams from manufacturing systems. This type of cloud-based manufacturing benefits from open innovation and cross-continent physically isolated product manufacturing. 

\subsection{Industrial Time Series Modeling}
The achievement of zero-defect in SFS is a major challenge. In SFS, all manufacturing components are perceived to be highly connected to ensure high-quality production. The term zero-defect refers to ensuring high-quality production during the execution of a complete manufacturing process~\cite{cheng2016industry}. To this end, industrial time series modelling ensures the proper monitoring of all manufacturing components during operations. However, data collection from multiple components results in high-dimensional data streams. The neo-fuzzy neuron (NFN) time series modelling method is adopted by IIoT systems. NFN can collaboratively connect the input data streams with the final outputs. NFN benefits from the convergence of input data, which results in decreased data streams and thus less iteration for learning model generation~\cite{zurita2016industrial}.

\subsection{Intelligent Shop Floor Monitoring}
The term physical Internet (PI) refers to the integration of cloud manufacturing with wireless and networking technologies. PI in IIoT systems provides the backbone to IIoTs and smart manufacturing object tracking systems based on radio-frequency identification. These smart manufacturing objects represent different forms of products during manufacturing after each process~\cite{zhong2017big}. However, IIoT systems need to track these smart objects during production to ensure that analytic processes provide intelligent shop monitoring. Researchers have proposed a BDA-based approach for the trajectory clustering of moving objects in shop floors. Although initial findings have been previously presented, a component-based BDA architecture is still necessary to develop highly optimised and intelligent smart object tracking systems for shop floor monitoring~\cite{zhong2017big}. Performance analysis and exception diagnosis model have been proposed and tested using Petri nets and decision tree algorithms~\cite{zhong2017big}. The model shows feasibility, and its real implementation in IIoT systems may help correctly quantify the results.

\subsection{Industrial Microgrids}
Massive data production in IIoT systems is evident due to feature-rich sensory and large-scale deployment of IIoTs in SFS~\cite{gamarra2016knowledge}. Therefore, manufacturing and environmental data, along with energy consumption data, can lead towards optimised energy utilisation in SFS. The application of BDA processes on these data silos can help improve planning, managing and utilising energy. Researchers have proposed BDA analytics methods for industrial-level microgrid planning in SFS. However, quantifiable studies that can lead towards efficient microgrid planning in IIoT systems are still required~\cite{gamarra2016knowledge}. 

\subsection{Monitoring Machine Health}
Prognostic health monitoring (PHM) helps find the machine behaviour for value creation during mechanical operations and facilitate machine data collection and management for the early diagnoses and prediction of machine faults. Several studies have performed analysis of PHM data~\cite{fleischmann2016improving,nunez2017ontology,nabati2017data}. In accordance with multiple International Standards Organisation and International Electrotechnical Commission and Society of Automotive Engineering standards, the authors of~\cite{nunez2017ontology} analysed ontological models developed from PHM data. These ontological models represent the hierarchical and semantic relationships amongst different machine components. The remaining useful life of machine components, faults, errors and failures during machine operations has also been explored. Studies have also presented dependency and failure mode analyses of different machine components. The analysis of PHM data helps plan and schedule machinery maintenance activities, thereby supporting in finding maintainable machine components before total failure. However, finding the relationship amongst different attributes and the failure impact of understudied machine components on other components in large-scale manufacturing environments is a challenging task~\cite{nabati2017data}. 

\subsection{ Intelligent Predictive and Preventive Maintenance} 
Predictive and preventive maintenance are the key requirements of large-scale IIoT systems~\cite{wang2016intelligent}. The BDA process can help in off-line prediction (\textit{i.e.,} performing prediction on the basis of historical data) and online maintenance (\textit{i.e.,} maintaining machines without shutting down the manufacturing units). Researchers have integrated Hadoop and Storm technologies for big data processing and used neural network-based methods for prediction~\cite{wan2017manufacturing}. The concept of adopting BDA for intelligent predictive maintenance is novel. However, new avenues need to be explored to fully realise a real-time prediction system.

\section{Taxonomy}
\label{Taxo}
\begin{figure*}[!b]
	\centering
	\includegraphics[width=\textwidth]{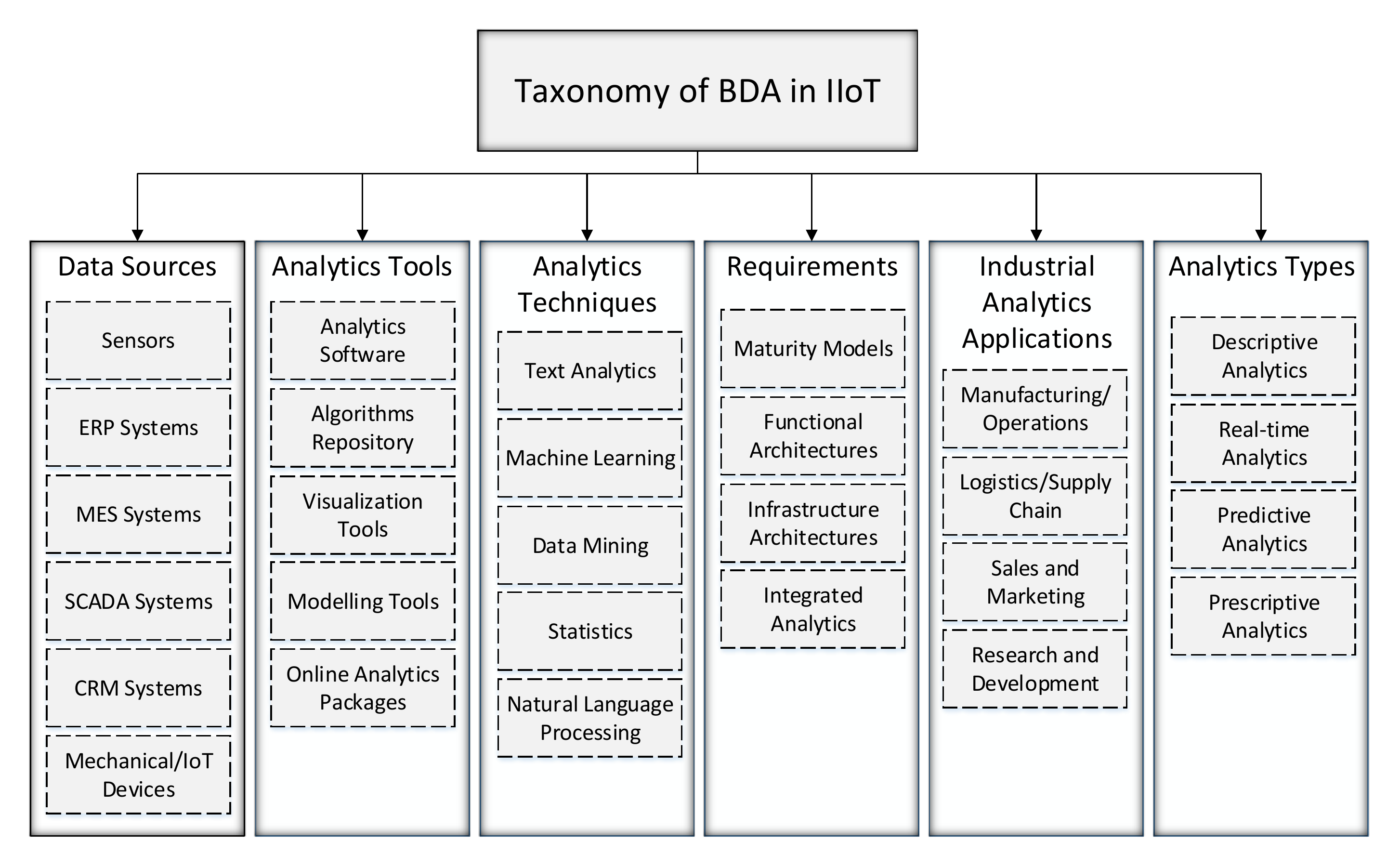}
	\caption{Taxonomy of BDA in IIoT}
	\label{fig:taxonomyofbigdataanalyticsinIIoT}
\end{figure*}

Figure~\ref{fig:taxonomyofbigdataanalyticsinIIoT} presents the taxonomy that is devised on the basis of data sources, analytics tools, analytics techniques, requirements, industrial analytics applications and analytics types.

\subsection{Data Sources}
In an industrial environment, numerous sources of data production, such as sensors, enterprise resource planning (ERP) systems, manufacturing execution systems (MES), supervisory control and data acquisition (SCADA) systems, customer relationship management (CRM) systems and machine/IoT devices. ERP systems enable organisations to employ a system that is composed of multiple integrated applications for managing business needs and automating many back-office functions related to technology, services and human resources. MES helps keep the track record of all manufacturing information in real time and receive up-to-date data from robots, machine and IoT devices \cite{ARIYALURANHABEEB2018}. SCADA systems are used to monitor and control a plant or equipment in industries (\textit{e.g.} telecommunications, water and waste control, energy, oil and gas refining and transportation). CRM systems are commonly used to manage a business–customer relationship. Machines and IoT devices are also deployed in industries to perform specific tasks, which generate an enormous amount of data on a daily basis. Applying analytics solutions to the collected data through all the above-mentioned systems, machines and IoT devices can extract valuable information that can help in decision-making purposes.

\subsection{Analytics Tools}
Several analytics tools are required to gain insights into a large amount of industrial data. These tools include analytics software, algorithm repository, visualisation tools, modelling tools and online analytics packages. Analytics software helps make predictions about unknown events. An algorithm repository is a crowd-sourced repository of algorithms that is designed by analysts using a common set of languages and a common interface. Visualisation tools help present data in advanced formats (e.g. infographics, dials and gauges, geographic maps, sparklines, heat maps and detailed bar, pie and fever charts). Modelling tools are used to define and analyse data requirements for supporting business processes within the scope of corresponding information systems in industries. Online analytics packages help keep track of and analyse data about web traffic.
\begin{landscape}

\begin{table*}[!t]
	\renewcommand{\arraystretch}{1}
	\caption{BDA Implementations in IIoT Systems}
	\label{tabletechnologies}
	\centering
    \small
	\begin{tabular}{|cp{30mm}p{20mm}p{25mm}p{20mm}p{35mm}p{35mm}p{30mm}|}
		\hline
		\rowcolor{honey2}
		\textbf{Ref.} & \textbf{Problem(s)}&\textbf{Objective(s)}&\textbf{Analytic Component(s)}&\textbf{Mode}&\textbf{Strengths}&\textbf{Limitations}&\textbf{Potential Solutions}\\
		\hline
		\rowcolor{Gray}
		~\cite{saldivar2016identifying}& Finding accurate customers' attributes for mass customization.& Self-prediction& Genetic Algorithm \newline k-means clustering& Historical data& Smart product development \newline Rapid response to customer needs & Needs to be implemented for real-time, Lacks large-scale validation in BDA environments & Deep Learning for BDA\\
		\hline
		\rowcolor{honey1}
		~\cite{saldivar2016self}& Finding accurate customers' attributes for mass customization.& Self-prediction& Self-organizing map& Historical data& Smart product development \newline Rapid response to customer needs & Needs to be implemented for real-time, Lacks large-scale validation in BDA & Re-enforcement learning algorithms\\
		\hline
		\rowcolor{Gray}
		~\cite{yang2017towards}& Enabling product customization and personalisation & Self-configuration \newline Self-organization& Highlighted, but no real implementation discussed & Streaming data& An end-to-end model for massive production and personalisation &No real implementation & Use-case implementation\\
		\hline
		\rowcolor{honey1}
		~\cite{zurita2016industrial}& Achieving zero-defect problem& Self-configuration& Neo-Fuzzy Neuron& Batch data& Performs industrial process monitoring and modelling & Accuracy needs to be implemented & Using alternate ML algorithms\\
		\hline
		\rowcolor{Gray}
		~\cite{zhong2017big}& implementing Physical Internet concept in manufacturing shop floors& Self-prediction& Decision trees& Batch data& The implementation results in better prediction rate& Performance values for different workers needs to be well-defined to control the rate of overestimation & Using alternate ML algorithms\\
		\hline
		\rowcolor{honey1}
		~\cite{gamarra2016knowledge}& Developing a proactive and sustainable microgrid& Self-prediction& A generic framework for knowledge discovery& Batch Data& An end-to-end approach for microgrid data analysis& Efforts are needed to explore analytics for full value chain level knowledge discovery in industrial microgrids & BDA Platform for full value chain Analytics\\
		\hline
		\rowcolor{Gray}
		~\cite{wan2017manufacturing}& Active preventive maintenance& Self-maintenance& Neural Networks& Batch data& Real-time active maintenance & Need to be investigated with real-time streaming data & Real-time BDA platform\\
		\hline
	\end{tabular}
\end{table*}

\end{landscape}

\subsection{Analytics Techniques}
Various analytics techniques that can help obtain value from big industrial data are available, thereby leading to making fast and better decisions. These analytics techniques include text analytics, machine learning, data mining and statistical and natural language processing (NLP) techniques. Text analytics helps derive high-quality information by unveiling patterns and trends using statistical pattern learning. Machine learning techniques enable industrial devices and machines to enter into a self-learning mode without being explicitly programmed. Data mining solutions enable enterprises to transform raw data into knowledge. Statistical tools help collect, summarise, analyse and interpret large amounts of industrial data, which lead to knowledge discovery. In an industrial environment, NLP tools are used to extract and analyse unstructured data.

\subsection{Requirements}
Certain requirements should be incorporated whilst developing new analytics systems for IIoT. These requirements include maturity models, functional architecture, infrastructure architecture and integrated analysis. Maturity models help measure and monitor the capabilities of analytics systems. They also help measure the effort required to complete a specific development stage. In summary, these models help monitor the health of an organisation’s big data programs. Functional architecture is an architectural model that helps identify the functions of analytics systems and their interactions. In addition, it defines how system functions work together to perform a specified system mission. In an industrial environment, analytics systems must be developed such that they can handle an enormous amount of data in real time. In this context, big data infrastructure requires experienced scientists to design the infrastructure from existing equipment in an industrial paradigm. One of the key requirements for analytics systems is that they should support the integrated analysis of multiple types of industrial IoT data.

\subsection{Industrial Analytics Applications}
Typical industrial analytics applications across the industrial value chain are as follows: manufacturing/operations, logistics/supply chain, marketing/sales and research and development. The use of predictive analytics in manufacturing can lead to rescheduling a maintenance plan prior to machine failure by considering past machine performance history. Moreover, it can help develop decision support systems for industrial processes. The appropriate use of analytics can play an important role in the logistics/supply chain (\textit{e.g.} condition monitoring, supply chain optimisation, fleet management and strategic supplier management). Analytics can help identify failing parts during product usage through sensor readings and gradually improve product characteristics (research and development). In the marketing field, analytics tools enable enterprises to predict and enhance future sales (\textit{e.g.} help in determining seasonal trends that can lead to developing an adaptive marketing strategy).

\subsection{Analytics Types}
Analytics has four types: descriptive, real-time, predictive and prescriptive analytics. Descriptive analytics helps gain insights into historical data (\textit{e.g.} number of defective items in the past and the reason for the defects). Meanwhile, real-time analytics enables enterprises to become aware of current situations (\textit{e.g.}, current status and location of a product and detection of a faulty machine). By contrast, predictive analytics helps identify potential issues that can occur in advance by using statistical and machine learning techniques (\textit{e.g.} expected inventory levels, anticipated demand levels, and prediction of equipment failure). Lastly, prescriptive analytics provides advice or suggestion on the best possible action that an end user should take (\textit{e.g.} whether a machine is receiving the right raw materials in the correct amount).

\section{Frameworks and Case Studies}
\label{frameworkscase}
Value creation is a major sustainability factor in modern enterprises whereby BDA processes are becoming the primary driver in creating values for customers and enterprises~\cite{Rehman2016}. IIoT systems are no exception. BDA processes can facilitate the amalgamation of customer and enterprise data to ensure massively customised production with zero defects. IIoT systems essentially integrate historical and real-time stored and streaming data at various levels. This multisource data integration leads to highly effective designs for new business models. Enterprises focus on different aspects of industry-wide value creation mechanisms, such as defining value propositions, value capturing mechanisms, value networks and value communication strategies for internal and external stakeholders~\cite{ehret2017unlocking}. Ideally, BDA processes can facilitate enterprise-level value creation whereby inbound intelligence is obtained by creating value for internal enterprise operations. Alternatively, outbound intelligence leads towards value creation for customers. Despite these opportunities, unlocking the perceived value from BDA technologies is challenging. The existing literature presents only a few such frameworks and use cases as follows. 

\subsection{SnappyData}
SnappyData is an open-source BDA framework that integrates Apache’s Spark and GemFire technologies~\cite{mozafari2016snappydata}. Apache’s Spark is adopted for big data processing, whereas GemFire facilitates highly scalable in-memory transactional data storage. The strength of SnappyData is its unified BDA engine that facilitates the performance of different types of analytical operation, such as online transaction processing, online analytical processing and streaming the data analytics of operational data. Despite its high performance, SnappyData still underperforms in cases with highly streaming data, which causes a bottleneck in real-time interactive visualisation performance.

\subsection{Ipanera}
Soilless food production systems, such as Ipanera, are being aligned with IIoT systems~\cite{de2016ipanera}. Ipanera continuously monitors water level and fertilizer quality in a field and generates insights for self-configuration. Although researchers have presented the concept, the Ipanera architecture involves multiple layers of physical devices and systems. It includes sensor nodes at the end points that actively collect data streams and transfer them to nearby IIoT clusters. These clusters are responsible for end point management, communication and configuration in a field. In addition, IIoT clusters provide feedback to end points to reconfigure their data collection behaviour. IIoT clusters transfer data streams to distributed analytics servers that run Apache’s Hadoop~\cite{vavilapalli2013apache}, MapReduce~\cite{dean2008mapreduce} and Spark~\cite{meng2016mllib} technologies for data processing and BDA. Ipanera provides support for streaming analytics that is used to trigger alerts for end points in case a new event is detected. Persistent storage and on-the-air configurations are two innovative features of the Ipanera architecture. This architecture is currently under development; hence, the complete design of the proposed architecture is still unavailable.

\subsection{Fault Detection Classification}
Large-scale distributed cyber manufacturing systems are based on multiple interconnected but geographically dispersed manufacturing units~\cite{lee2017framework}. The fault detection and classification (FDC) framework finds manufacturing faults in products. The core of the FDC architecture is the integration of IoT devices into CPS and cloud computing technologies. IoT devices in production facilities continuously collect and analyse data streams to detect various signals that are transferred to back-end cloud servers. These cloud servers execute BDA processes to detect and classify faulty products using deep belief networks based on deep learning methods~\cite{lee2009convolutional,hinton2006fast}. FDC was analysed by deploying it in a car headlight manufacturing unit that produced reliable results. 

\subsection{BDA Architecture for Cleaner Production}
The term cleaner production refers to ensuring reduced environmental impacts during the execution of the entire product life cycle. It is based on three phases~\cite{zhang2017big}. The first phase is about product design and manufacturing. The second phase involves product use, service provisioning and maintenance. The third phase is concerned with product remanufacturing, reuse and recycling. Considering the importance of such clean technologies, researchers have proposed a four-stage BDA architecture. In the first stage, the architecture considers value creation objectives during a product’s life cycle, such as improving product designs and ensuring energy efficiency, proactive maintenance and environmental efficiency. In the second stage, big data acquisition and integration are performed using IoT devices. In the third stage, big data are processed using Apache’s Hadoop and Storm technologies. Finally, BDA processes are executed in the fourth stage whereby the architecture provides clustering, classification, association rule mining and prediction-related algorithms. The proposed architecture was evaluated and tested on an axial compressor manufacturing unit. The annual reports of the production unit show that the proposed architecture realises all the value creation objectives for cleaner production.

\subsection{Smart Maintenance Initiative: Railway Case Study}
Apart from SFS, Japan is attempting to upgrade its railway system to a new level by adopting IIoT systems for the smart maintenance of railway tracks~\cite{takikawa2016innovation}. To achieve its ‘smart maintenance vision’, Japan’s railway is adopting IIoT, BDA and automation technologies. The ‘smart maintenance vision’ will provide a solution to four challenges:  1) ensuring condition-based maintenance, 2) providing work support through artificial intelligence (AI), 3) managing railway assets and 4) performing database integration. The progress details of Japan’s railway towards this vision are available in this report~\cite{takikawa2016innovation} for interested readers. 

\section{Opportunities, Research Challenges, and Future Technologies}
\label{future}
Considering the vision of IIoT systems, BDA will evidently help enterprises in the value creation process. BDA processes will maximise operational efficiency, reduce product development cost, ensure massively customised production and streamline the supply chain management. However, this review shows that the existing literature is considerably lagging behind this vision. Table~\ref{tablechallenges} presents the summary of research challenges and their perceived solutions to fully adopt IIoT systems in BDA.

\subsection{Opportunities}
The adoption of BDA processes in IIoT systems results in multidimensional research opportunities.

\subsubsection{Automation and AI}
The enrichment of intelligent features can lead towards highly optimised and automated industrial processes~\cite{cao2016big,dopico2016vision}. Therefore, AI will be the core component of big data optimisation and analytics, which will result in highly efficient industrial processes~\cite{wang2016towards}. Future IIoT systems will integrate and ingest big data from various online and off-line and inbound and outbound operations. The integration of customer and enterprise data will result in high-dimensional, multi-million variable datasets. AI methods will help optimise and analyse such big datasets~\cite{berger2016research,rehman2016big}.

\subsubsection{Human Machine Interaction}
Wearable computing and augmented reality technologies are leading towards new human–machine interaction models and interfaces~\cite{toniges2017emerging,oks2017application}. The enrichment of such interaction models with real-time knowledge patterns from big data systems will result in highly productive and rich user interfaces. In addition, robotics technologies (for physical and virtual robots) will be widely adopted by future IIoT systems. Therefore, BDA processes will enrich intelligence to produce highly autonomous and self-sustaining non-obtrusive systems.

\subsubsection{Cybersecurity, Privacy, and Ethics}
Cybersecurity will become an essential requirement due to connected intelligence in IIoT systems. BDA processes will help provide real-time cyber threat intelligence by analysing security attacks, privacy leaks, unauthorised data access and unethical data collection~\cite{martellini2017assessing}. In addition, BDA processes will help analyse network and information security-related enterprise data to find anomalies, outliers, threats, attacks and vulnerabilities across IIoT systems~\cite{jensen2017big}.

\subsubsection{Universal Standards}
The adoption of BDA processes is still in its initial stage; thus, existing systems may not be compliant with universal standards across all or multiple industries~\cite{wollschlaeger2017future,igor2016proposal}. New universal standards are required to define the type of big data that the industries can collect from customers, determine how data should be secured, preserved and shared and identify the stakeholders who will benefit from the data. In addition, standards must also ensure the perceived benefits to customers in exchange for their personal data. These universal standards will help address ethical issues in big data systems and create value for customers by providing personalised products and services. 

\subsubsection{Protocols for Interoperability}
Practically, multiple industries are involved in the entire process—from customer data acquisition to finished product/service and supply chain management~\cite{wollschlaeger2017future}. Interoperability is a major consideration among different industries; however, new protocols are required to realise fully interoperable IIoT systems. These protocols can lead towards value creation for enterprises, although a few questions must be addressed, such as what are the interoperability parameters, how will BDA processes be executed in cross-industry systems and how will heterogeneity in data, computing technologies and industrial production systems be handled. A well-defined interoperability protocol can help answer these questions.

\subsubsection{End-to-end Industrial Analytics}
Big data in IIoT systems evolves from multiple inbound and outbound data sources, such as customer data and operational data from finance, marketing, human resources, IoT devices, CPS and manufacturing systems~\cite{dagnino2014industrial}. Nevertheless, existing systems manage all these data sources separately to execute BDA processes. An opportunity exists to develop an end-to-end industrial analytics pipeline that can handle big data from various data sources in parallel and find highly correlated knowledge patterns that emerge across entire IIoT systems \cite{de2019multi}.

\subsubsection{Precision Manufacturing}
BDA processes can help enrich precision manufacturing systems~\cite{wu2017omic}. The classification and categorisation of customers’ needs and behaviour-related data can lead towards innovative product designs. Enterprises will be able to offer the right products and services to the right customers. Precision manufacturing will considerably help in equal value creation for customers and enterprises. Early examples of precision manufacturing systems are available in the healthcare industry \cite{wan2019reconfigurable}. However, these systems should be integrated into IIoT systems~\cite{wu2017omic}.

\subsection{Research Challenges and Future Technologies}
Considering the opportunities, research efforts are required to improve the entire technology ecosystem for IIoT systems. 
\subsubsection{Big Data Process Integration into IIoT Systems}
Ideally, IIoT systems should execute real-time highly interactive big data applications. In practice, however, considerable effort is required for planning, creating, deploying, maintaining and continuously improving domain-specific big data processes for each industry. Future BDA processes should be able to provide real-time knowledge patterns and industry-wide intelligence through single dashboard applications. In this regard, all legacy and state-of-the art data sources should be vertically aligned such that enterprises can easily analyse and correlate different industrial processes and operations.

\subsubsection{Orchestrating BDA Applications Using Concentric Computing}
Concentric computing systems provide computational and storage support through different devices and systems~\cite{concentric2017}. Thus, massive heterogeneity should be addressed in terms of processing capabilities, in-memory and disk-based storage systems, battery-powered and fully powered devices and systems and multiple communication channels with varying bandwidth capacities~\cite{ur2016towards}. Big data applications on top of concentric computing systems should be designed by considering efficiency objectives in terms of storage, in-network data movement, energy consumption, privacy, security and real-time knowledge availability~\cite{YAQOOB2018, yaqoob2017rise}. In this regard, priority should be given to devices and systems near data sources. This approach can help maximise value creation for enterprises in terms of operating cost for big data systems. Given that maximum data collection, filtration and processing are performed before data arrive in cloud computing systems, the operational costs for data storage and cloud service utilisation will therefore be minimised~\cite{Rehman2016}. Another benefit of concentric computing systems is their ability to ensure real-time or near real-time intelligence near end points, IoT devices and other data sources in IIoT systems~\cite{georgakopoulos2016internet}. 

\subsubsection{Emerging and Complimentary Technologies for IIoT Systems}
On the one hand, BDA adoption is increasing in IIoT systems. On the other hand, IIoT systems should address massive heterogeneity without compromising overall operational efficiency due to emerging, complementary technologies, such as IoT. Considering this condition, a few technologies will become integral parts of future IIoT systems. 

Virtualisation is the essence of distributed systems, such as cloud computing systems and concentric computing systems. Virtualisation is traditionally performed at multiple levels, such as operating systems, networks, storage, applications and hardware. Operating system-level virtualisation is the most common whereby operating system kernels and functions are virtualised as virtual machines (VMs). However, the mobility of IoT devices requires continuous VM migration among different computer servers~\cite{wollschlaeger2017future,ma2017sdn}. Containerisation is the emerging technology that is gradually replacing VMs by sharing a single kernel among different applications on the same type of operating systems. Containerisation technologies offers more secure and faster processing; hence, they have become highly beneficial for addressing timeliness and latency issues when BDA processes are executed using VMs~\cite{gupta2017evolution}. 

Large enterprises traditionally adopt highly coupled SOAs, which are difficult to test and result in high maintenance cost. Microservices are emerging alternatives to SOAs whereby highly scalable and loosely coupled cloud services are orchestrated~\cite{dang2017graph}. The microservice architecture can be adopted best for BDA processes because these processes should be executed across multiple platforms and devices in IIoT systems~\cite{fokaefs2016enabling}. The details of microservice architecture’s implementation are available in~\cite{wagner2016microservices} for interested readers.

The multipoint, multisite and high-dimensional data production in IIoT systems results in complex big datasets. Graph and network theories can help reduce this massive complexity~\cite{jacobs2016large}. Graph data structures and big graph analytics methods can be adopted to separate, map and analyse big data in different graph formats. The adoption of big graph analytics can lead towards efficient and highly optimised execution of BDA processes across IIoT systems.

\begin{landscape}
\newcolumntype{g}{>{\columncolor{Gray}}p}
\newcolumntype{a}{>{\columncolor{honey1}}p}
\newcolumntype{d}{>{\columncolor{honey2}}p}
\begin{table*}[!t]
	\renewcommand{\arraystretch}{1}
	\caption{Summary of Research Challenges and their Perceived Solutions}
	\label{tablechallenges}
	\centering
    \small

	\begin{tabular}{|d{35mm} a{55mm} g{65mm} a{65mm}|}
		\hline
	
		\textbf{Type} & \textbf{Issues} &\textbf{Causes}&\textbf{Solutions}\\
		\hline
		Cybersecurity & - Internal Attacks \newline - External Attacks & - Security Vulnerabilities \newline - Openness of Systems& - Intelligent Monitoring Tools Needed\newline - Deployment of End-to-End Security Models is Essential \newline - System-wide Forensic Analysis should be performed periodically\\\hline
		Privacy& - Identity Breaches \newline - Personal Data Theft\newline - Business Data Leakage  & - Bad Security Models \newline - Absence of Standard Operating Procedures \newline - Weak Data and Information Sharing Policies& - Using Data Anonymisation Protocols \newline - Privacy preserving interaction models for users, devices, and systems \\\hline
		Big Data Processing& - Bad Data Integration \newline - Missing Data Streams \newline - High Latency& Heterogeneous Data Sources \newline - Mobility and Connectivity Issues \newline - Data overloading and Bandwidth Limitations& - Intelligent Real time Data Fusion\newline - Device-centric big data processing architectures\newline - Concentric Computing Models\\\hline
		
		Standardization& - Difficulty in Interoperability and and System Integration& - Absence of Global Standardization Body& - Developing Local, Regional, Industry-specific, and Global Standards\\\hline
	
		Connectivity and Communication& - Bad and Inaccurate Data Transfer \newline - Data Loss \newline - High Latency& - High Mobility\newline - Large Data Streams\newline - Congestion& - Need to create always-on, ultra-high available and reliable communication protocols \\\hline
		
		Scalability & - Resource Discovery \newline - Data offloading \newline - Data Management & - Low Processing Power at device-end \newline - Massive Data Production \newline - Realtime Actuation& - Near-device data processing, In-memory Data Processing, Edge Computing\\\hline
		System management & - Difficult to deploy, configure, monitor, and control large scale IIoT networks& - Cloud-centric & - Device-centric \\\hline
		Efficiency & - High Energy Utilization\newline - Resources-constraints \newline - Device-overloading& - Always-on IIoT Devices and Systems \newline - Massive and Continuous Data Generation and Device Operations \newline - On-device Data Management and Analytics & - Enabling Energy, memory, and computation-efficient algorithms and processes for big data processing, management and analytics in IIOTs \\\hline
	\end{tabular}

\end{table*}
\end{landscape}
Emerging technologies, such as fog computing and blockchain, can play a pivotal role in BDA for IIoT~\cite{8598784}. Fog computing has been widely used in IoT devices \cite{hassan2018role}, particularly those for IIoT and smart manufacturing, for localised and timely data processing and storage, and primarily to offset long delays that can be incurred in a cloud environment~\cite{el2017efficient}. Blockchain is the underlying technology for bitcoins; however, it has been foreseen as a distributed ledger that can provide decentralised storage for data generated by IoT devices. Data are stored in a blockchain ledger with high integrity, authenticity, resiliency and trust~\cite{almadhoun2018}. All transactions are cryptographically signed by IIoT devices and validated in a decentralised manner without an intermediary. The data origin is validated before being recorded on the ledger. Moreover, blockchain smart contracts can be used to provide decentralised authentication, management and control access to data generated by IIoT devices. Smart contracts are basically codes that are executed by all blockchain miners, and the execution outcome is verified and agreed upon by all mining nodes. Furthermore, given the limited computing, networking and storage capacities of IIoT devices, fog nodes are envisioned to be equipped with cloud and blockchain interfaces in the future to communicate and interface with the cloud environment and the blockchain network~\cite{almadhoun2018}. 

\section{Conclusions}
\label{conclusions}
The vision of Industry 4.0 to connect manufacturing systems with distributors and consumers can only be achieved by adopting IIoT and BDA processes as core components for value creation. This paper discusses the rise of big data in IIoT systems and presents a detailed survey of related technologies, algorithms, frameworks and case studies. A detailed taxonomy is provided to classify the key concepts in this important research area. Several indispensable frameworks and case studies are outlined and discussed. Furthermore, we present a detailed discussion of future opportunities, technologies and research challenges. 
We conclude that the adoption of BDA in IIoT systems is still in its early stage. Research on complementary components of IIoT systems, such as IoT devices, augmented reality and CPS, is also in its infancy. Current BDA systems provide generic frameworks for data engineering, preparation and analysis. However, considerable effort is required to alter existing BDA processes to meet the demands of IIoT systems. Future research should be conducted to devise new standards for interoperability among cross-Industry 4.0 BDA platforms and to provide capability for end-to-end reliable application processing by considering the anatomy of concentric computing systems.

\section*{Acknowledgement}

Imran's work is supported by the Deanship of Scientific Research at King Saud University through Research group No. (RG \# 1435-051). 

\section*{References}
\bibliographystyle{model1-num-names}
\bibliography{sample}

\end{document}